\documentclass[aps,pra,twocolumn,groupedaddress,superscriptaddress,showpacs]{revtex4}

\usepackage{graphicx}
\usepackage{amsmath,amssymb,amsfonts}
\usepackage{natbib}

\newcommand{\be}{\begin{equation}}
\newcommand{\ba}{\begin{eqnarray}}
\newcommand{\ee}{\end{equation}}
\newcommand{\ea}{\end{eqnarray}}
\begin{document}

\title{Bohmian trajectory analysis of high-order harmonic generation: ensemble averages, non-locality and quantitative aspects}

\author{J. Wu}
\affiliation{Department of Physics and Astronomy, University College
London, Gower Street, London WC1E 6BT, United Kingdom}

\author{B. B. Augstein}
\affiliation{Department of Physics and Astronomy, University College
London, Gower Street, London WC1E 6BT, United Kingdom}
\affiliation{School of Chemistry, University of Leeds, Leeds LS2 9JT, United Kingdom}

\author{C. Figueira de Morisson Faria}
\affiliation{Department of Physics and Astronomy, University College
London, Gower Street, London WC1E 6BT, United Kingdom}

\date{\today}

\begin{abstract}
We perform a Bohmian-trajectory analysis of high-order harmonic generation (HHG), focusing on the fact that typical HHG spectra are best reproduced by the Bohmian trajectory starting at the innermost part of the core [Phys. Rev. A \textbf{88}, 023415 (2013)]. Using ensemble averages around this central trajectory, we show that, for the high-plateau and cutoff harmonics, small ensembles of Bohmian trajectories are sufficient for a quantitative agreement with the numerical solution of the time-dependent Schr\"odinger equation (TDSE), while larger ensembles are necessary in the low-plateau region. Furthermore, we relate the Bohmian trajectories to the ``short" and ``long" trajectories encountered in the Strong-Field Approximation (SFA), and show that the time-frequency maps from the central Bohmian trajectory overestimate the contributions of the long SFA trajectory, in comparison to the outcome of the TDSE computations. We also discuss how the time-frequency profile of the central trajectory may be influenced nonlocally by degrading the wave-packet propagation far from the core.
\end{abstract}


\maketitle



\section{Introduction}

Since its discovery in the late 1980s, high-order harmonic generation (HHG) has attracted
a great deal of attention due to the myriad of possible
applications, such as table-top, extreme ultraviolet (XUV) sources
\cite{waterwindow}, high-frequency light pulses of attosecond
duration \cite{attopulses,Hentschel_2001} (for a review, see
Ref.~\cite{altucci:JModOpt:2012}), and the attosecond
imaging of dynamic processes in matter
\cite{Kienberger_2004,Itatani_2004,Goulielmakis_2008,Smirnova_2009,Vozzi_2011}.
These applications have been made possible due to a very intuitive
physical interpretation of HHG in terms of electron orbits, in which
an electron, under the influence of the external laser field,
reaches the continuum by tunnel ionization, is accelerated by the
field, and, at a subsequent time, recollides with its parent ion
\cite{corkum:PRL:1993}. Upon recollision, the electron may recombine
with a bound state, thus emitting high-frequency radiation. This physical picture is widely known as the three-step model (TSM).

In early HHG studies, this picture has been put across using an ensemble of classical electrons being released in the continuum at different times within a field cycle and returning to the core \cite{corkum:PRL:1993}. These classical computations were hugely successful in explaining typical features observed in high-order harmonic spectra, namely a plateau consisting
of harmonics of comparable intensities, followed by a sharp decrease in the harmonic signal, the so-called cutoff, whose energy position is proportional to the driving-field intensity.

Quantum mechanically, the TSM has been extracted from the phase of the time-dependent electronic wavefunction. The most widely known example is the steepest descent method applied to the expectation value of the dipole operator, within the framework of the strong-field approximation (SFA) \cite{lewenstein:PRA:1994}. In the SFA, two key assumptions are made: (i) the continuum is approximated by field-dressed plane waves; (ii) the core is reduced to a source term, located at the origin of the coordinate system, so that its internal structure, such as excited bound states, is neglected. Despite these simplifications, the SFA has been hugely successful, as far as qualitative predictions are concerned. Other, more recent approaches based on the recollision picture are the Volkov-eikonal approximation \cite{Olga2006,Olga2008}, the Coulomb corrected Strong-field approximation \cite{Bauer2008,Bauer1,Bauer2}, the Herman Kluk propagator \cite{vandeSand1999,Zagoya2012}, the adiabatic approximation \cite{adiabatic1,adiabatic2} and the coupled coherent states method \cite{CCS2012}. These approaches go beyond the SFA as the Coulomb potential is incorporated in the electron propagation, even if in many cases approximately.
Furthermore, the TSM has
also been inferred from the numerical solution of the time-dependent Schr\"{o}dinger equation
(TDSE), by time-frequency analysis (see, e.g., \cite{timefrequency2,FDS1997,Belgium1998,timefrequency1}; for recent articles see \cite{Ruggenthaler_2008,Chirila_2010,Ciappina_1_2012,Ciappina_2_2012}).

The success of orbit-based quantum mechanical models also implies that HHG may be related to the overlap between
the bound and continuum parts of the time-dependent wavefunction, which are coupled via the dipole operator.
Spatially, this overlap takes place near the core, and exhibits high-frequency oscillations that lead to
the plateau and cutoff. These oscillations have been first identified in the late 1990s \cite{Protopapas1996}
and have been studied in recent publications employing semiclassical propagators \cite{vandeSand1999,Zagoya2012,adiabatic1}.
Thereby, a necessary condition for obtaining a clear plateau and cutoff is spatial localization \cite{Protopapas1996}. Consequently, the acceleration form of the dipole operator, which probes regions near the core,
leads to better-quality spectra than the dipole length, which emphasizes large distances
 \cite{burnett:PRA:1992,Krause:PRA:1992}. This happens even if, in principle, both formulations are equivalent
 \footnote{This is a consequence of the Ehrenfest theorem, as shown in \cite{burnett:PRA:1992}.
   For approximate methods, such as the SFA, however, the Ehrenfest theorem may become
   invalid (for a systematic study see Ref.~\cite{Granados_2012} and for specific examples
   in molecular systems see Refs.~\cite{Chirila_2007,Augstein_JMO_2011})}.

In order to understand this overlap in detail, it is helpful to study the probability-density flow associated
with the quantum mechanical wavefunction in specific configuration-space regions, without losing phase information. This can be performed using Bohmian trajectories, which act as ``tracer particles'' \cite{bohm:PR:1952-1,sanz-bk} (for strong-field applications see
 \cite{lai:EPJD:2009,botheron:PRA-2:2010,takemoto:JCP:2011,oriols-bk,botheron:PRA-1:2010,Song_2012,Wu2013}). Bohmian trajectories, however, are highly non-local, non-classical entities. Indeed, a spatially localized Bohmian trajectory may contain bound and continuum dynamics, while, classically,
  one may identify a bound or unbound trajectory by looking at the spatial regions it occupies \footnote{Bohmian trajectories can only be related to classical
trajectories associated to the original potential under very specific conditions, namely for coherent
states and values of the Mandel parameter characterizing the quasi-Poissonian regime; for a detailed
discussion see S. Dey and A. Fring, Phys. Rev. A \textbf{88}, 022116 (2013).}. 

In a previous publication \cite{Wu2013}, we have found that the
spectrum obtained from the Bohmian trajectory located at the
innermost part of the core is qualitatively very similar to that of
the dipole acceleration. Furthermore, we have shown that the
time-frequency maps from the central Bohmian trajectory can be
associated to an ensemble of classical trajectories leaving and
returning to the core.  Physically, these results lead to a more
restrictive statement than to say that HHG takes place near the
core. In fact, they show that the main features predicted by the TSM
may be obtained even if the spatial extension of the core is
neglected. A natural question is, however, what regions of the the
core must be included, in order to obtain a quantitative agreement
between the TDSE and the Bohmian-trajectory computation. Clearly, as
the Bohmian trajectories are extracted from the TDSE, one expects
that eventually both methods will lead to identical outcomes.
However, is it sufficient to include the immediate vicinity of the
innermost trajectory, or should the whole core be considered?
Furthermore, it has been recently shown that short-range potentials
overestimate the contributions of the long TSM trajectories, as
compared to their long-range counterparts \cite{Chirila_2010}. This
has been attributed to the spatial extension of the core and to the
Coulomb tail. Do we find similar effects? If so, how can they be
understood in the present framework?

Another open issue is non-locality.
In Ref.~\cite{Wu2013}, we have briefly shown that we central Bohmian
trajectory is affected by the probability flow far from the core, and
 suggested that this is due to non-local transmission via the phase of the wavefunction. A more detailed study of this non-locality, and in particular of how this phase builds up, has not yet been performed. 

In the present paper, we address the above-mentioned questions. Our work is organized as follows.
 In Sec.~\ref{theory} we provide details about our model, including our TSDE and Bohmian trajectory computations. Subsequently, in Sec.~\ref{results}, our results are presented. In Sec.~\ref{ensemble}, using ensembles of Bohmian trajectories, we investigate different regions of the core, and how a quantitative agreement with the TDSE may be reached. In Sec. \ref{timefrequency}, we relate the Bohmian trajectories to those obtained from the SFA using time-frequency analysis, and show how the dynamics far from the core affects the central trajectory nonlocally via the phase of the wavefunction.  The main conclusions
and results are summarized in Sec.~\ref{conclusions}. Throughout, we employ atomic units.
\section{Theory}
\label{theory}
\subsection{Time-dependent Schr\"odinger equation}
For the sake of simplicity, we have developed our analysis in one
dimension, which contains the essential physical elements for
linear polarization. We solve the time-dependent Schr\"odinger equation
\be
 i\frac{\partial \Psi(x,t)}{\partial t} = H \Psi(x,t) ,
 \label{schro}
\ee
where $H=H_0+H_{int}(t)$ and $\Psi(x,t)$  denote the time-dependent Hamiltonian and wave function, respectively.
The field-free Hamiltonian $H_0$ is chosen as %
\be
 H_0 = -\frac{1}{2}\ \nabla^2 +V(x),
 \label{bare}
\ee
in which $V(x)$ is modeled by a long-range, soft-core potential.  The interaction with the field is $H_{int}(t)=+ x E_0 F(t)$, where the driving field is chosen to be a flat-top pulse of frequency $\omega_0$ (see our previous publication \cite{Wu2013} for details on both the pulse and $V(x)$).

The system is initially in the ground state of the field-free Hamiltonian $H_0$. This eigenstate has been numerically obtained by the imaginary time propagation method \cite{lehtovaara:JCompPhys:2007}, and has an energy $\epsilon_0=-0.66995$ a.u..
The exact time-propagation of the wavefunction according to
the TDSE (\ref{schro}) has been carried out by combining the
split-operator technique \cite{feit-fleck:JCompPhys:1982} with the
fast Fourier transform (FFT) technique \cite{press-bk-1}.

To ensure that all the relevant dynamics are incorporated for the
parameter range of interest, we have set the box boundaries located
far enough from the core region (at $l=150$~a.u.). Furthermore,
special care has been taken in order to avoid reflections and
spurious effects near the box edges. First, we have employed a mask
function in the form of \be
 \mathcal{M}(x) = \left\{ \begin{array}{lcc}
  \cos^{1/8}\left[\displaystyle \frac{\pi|x+x_1|}{2(-l+x_1)}\right]  , & \quad
   x \le -x_1 \\
  1 , & \quad
   x_1 \le x < x_1 \\
  \cos^{1/8}\left[\displaystyle \frac{\pi|x-x_1|}{2(l-x_1)}\right]
   , & \quad
   x \ge x_1
 \end{array} \right. ,
 \label{absorber}
\ee
which becomes active at $x_1$. This function is smooth enough, but
still capable to absorb with a high efficiency, thus avoiding
nonphysical reflections. We have tested this fact by considering a
wide range of box sizes. Second, the maximum/minimum value of the
total potential function has been truncated in order to avoid also
nonphysical accelerations towards the box edges. In this regard, the
size of the box was chosen such that this truncation takes place
close to the region where the absorber becomes active. Unless
otherwise stated, $x_1=145$ a.u. Furthermore, no filter functions (e.g., Hanning
windows) were used in order to avoid influencing the topology of the Bohmian trajectories \footnote{Hanning filters are commonly employed to eliminate the background that occurs when computing HHG spectra from the expectation value of the dipole length. They however force the probability-density flow to return to the core, and thus alter the Bohmian trajectories. This has been verified by constructing these trajectories from a spectrum in which this filter has been used.}.

The expectation value of the coordinate $x$ and the dipole acceleration are computed as
\begin{equation}
 \bar{x}(t)=\int^{+\infty}_{-\infty}dx\Psi^*(x,t)x\Psi(x,t)
 \label{length}
\end{equation}
and
\begin{equation}
 \bar{a}(t)=-\int^{+\infty}_{-\infty}dx\Psi^*(x,t)dV(x)/dx\Psi(x,t),
 \label{accel1}
\end{equation}
 respectively \cite{burnett:PRA:1992}. Note that $-\bar{x}(t)$ gives the expectation value of the dipole operator in its length form.

To avoid effects associated with the loss of
probability due to the presence of the absorber,
special care was taken in properly renormalizing the wavefunction when computing all expectation values in this paper. We have verified that this renormalization has a minor effect in $\bar{x}(t)$ and practically none in $\bar{a}(t)$.
\subsection{Bohmian trajectories}
\label{Bohmtraj}

In order to determine the Bohmian trajectories, first $\Psi(x,t)$ is recast in polar form,
\be
 \Psi(x,t) = \rho^{1/2}(x,t) e^{iS(x,t)} ,
 \label{psit}
\ee
where $\rho$ is the probability density and $S$ is the real-valued
phase. Substitution in the TDSE leads to two coupled differential equations in $\rho$ and $S$ \cite{Wu2013}. The Bohmian trajectories are obtained by integrating the equation of motion
\be
 \dot{x} =\nabla S  = \frac{1}{2i}
  \left(\frac{\Psi^* \nabla \Psi - \Psi \nabla \Psi^*}{|\Psi|^2}\right).
 \label{eom}
\ee
Eq.~(\ref{eom}) has been
integrated ``on the fly'', substituting the value of the
wavefunction $\Psi (x,t)$ at each time into the last term of this
equation. We have employed the Euler method, which has proven to be accurate enough.

The Bohmian versions of the expectation values (\ref{length}) and
(\ref{accel1}) are given by
\ba
 \bar{x}_B(t) & = & \frac{1}{N} \sum_{i=0}^N x_i(t) ,
 \label{dipoleB} \\
 \bar{a}_B(t) & = & -\frac{1}{N}
   \sum_{i=0}^N \frac{dV(x)}{dx} \bigg\arrowvert_{x=x_{i}(t)},
 \label{accel1B}
\ea
respectively. In these
expressions, $x_i(t)$ denotes the $i$th Bohmian trajectory from an
ensemble and $N$ is the total number of these trajectories in each
numerical experiment performed.

In all ensemble computations, we have considered a
set of initial conditions for the Bohmian trajectories
obtained from uniformly generated random
numbers within a certain interval $[-x_c,x_c]$, symmetric with
regard to the origin, $x=0$.  These random numbers are multiplied by
weights, which are chosen in such a way that the probability density
related to the ground state of the soft-core potential,
$|\Psi(x,0)|^2$, is mimicked.
Unless otherwise stated, we have chosen $x_c=4.102$~a.u. This ensures that
the time evolution of most of the probability density will be well
monitored, as the integrated probability for $|x| \ge x_c$ is only
0.151\% of the total probability.
\subsection{Fourier and Gabor transform}
The power spectra in this work are computed as
\be
 I(\omega) = \left\arrowvert \int g(t)\ \! e^{-i\omega t}
   dt \right\arrowvert^2=|a_F(\omega)|^2 ,
 \label{fourier0}
\ee
where $g(t)$ is a generic time-dependent function denoting either the expectation values discussed above or specific sets of Bohmian trajectories, and the integral $a_F(\omega)$ is the standard Fourier transform. We investigate the time-frequency profiles and the phase of the wavefunction using the Gabor transform
\begin{equation}
a_G(\omega, t)=\int dt^{\prime}g(t)\exp[-(t-t^\prime)^2/(2\sigma^2)]\exp(-i\omega t^{\prime}).
\label{Gabor}
\end{equation}
If $\sigma \rightarrow \infty$ the standard Fourier transform $a_F(\omega)$ is
recovered and all temporal information is lost. Here, we choose the
same temporal width as in Ref.~\cite{Chirila_2010}, i.e.,
$\sigma=1/(3\omega_0)$.

\section{Results and discussion}
\label{results}
\subsection{Harmonic spectra}
\label{ensemble}
\begin{figure}
 \begin{center}
 \includegraphics[width=8.5cm]{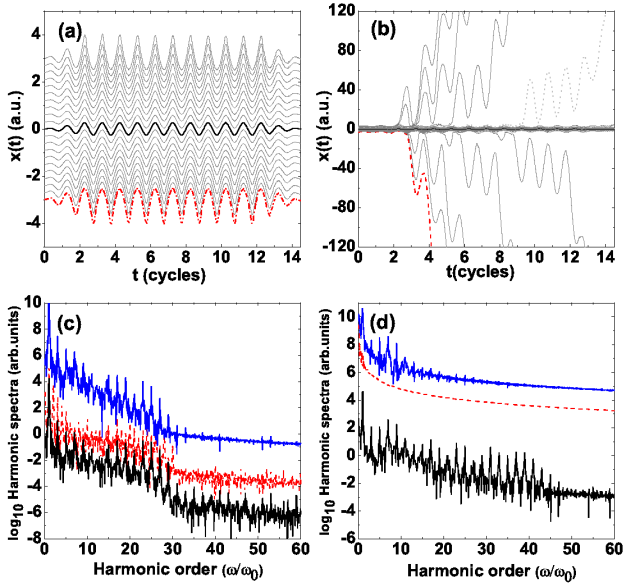}
 \vspace*{-3.4cm}
 \caption{\label{fig2} (Color online)
  Top panels: Bohmian trajectories obtained with a flat-top pulse with
  frequency $\omega_0=0.057$ a.u. and field amplitudes $E_0=0.05$ a.u. [panel (a)] and $E_0=0.075$~a.u [panel (b)].
Bottom panels: Power spectra computed from individual trajectories $x_i(t)$ selected
  from among the sets at the top panels, together with spectra from the whole ensemble average  (\ref{dipoleB}), for $E_0=0.05$ a.u. [panel (c)] and $E_0=0.075$~a.u [panel (d)]. The black solid line and the red dashed line correspond to the trajectories starting at $x(0)=0$~a.u. and $x(0)=-3$~a.u., respectively. To facilitate a direct comparison, we have used the same fonts to designate different types of trajectories as in Figs.~\ref{fig2}(a) and (b). The spectra computed from the ensemble average are depicted as the blue solid lines.
  To facilitate visualization, the spectra obtained from the ensembles have been shifted upwards in six orders of magnitude.  The corresponding driving-field intensities are $I=8.7 \times10^{13}\mathrm{W}/\mathrm{cm}^2$ (Keldysh parameter $\gamma=1.32$) and $I=1.97 \times 10^{14}\mathrm{W}/\mathrm{cm}^2$ (Keldysh parameter $\gamma=0.88$).}
 \end{center}
\end{figure}

For clarity, we will first discuss the behavior of individual Bohmian trajectories and their spectra. These results are displayed in Fig.~\ref{fig2}. The spectra from the central trajectory, starting at $x(0)=0$ exhibit a clear plateau and a cutoff at $\omega_c \approx
|\epsilon_0| + 3.17U_p$, where $U_p = E_0^2/4\omega_0^2$ is the
ponderomotive energy, for both moderate and high driving-field intensity. This is a very good example of the nonlocal behavior of a Bohmian trajectory. Classically, a trajectory confined in configuration space would imply bound dynamics. Quantum mechanically, however, what defines whether a system is bound or unbound is its energy. The cutoff in the spectra of the central Bohmian trajectory goes far beyond the ionization potential $|\epsilon_0|$. Hence, it contains bound and continuum dynamics.

For peripheral Bohmian trajectories, the situation is markedly different. As long as they perform an oscillatory motion near the core, the plateau and the cutoff will be present. If, however, they leave this spatial region, the corresponding spectra will only consist of the fundamental and of a uniform background. Moreover, the further from $x(0)=0$ the initial position of a trajectory is, the higher the overall intensity in the spectra will
be. Thus, even if the number of trajectories that leave the core is relatively small, their spectra are several
orders of magnitude more intense than those obtained from the
innermost trajectories. This masks the latter
contributions in the ensemble average (\ref{dipoleB}). This behavior is more extreme for higher intensities, as in this case the outward probability-density flow is larger [see blue solid lines in Figs.~\ref{fig2}(c) and (d)].

Formally, the ensemble average (\ref{dipoleB}) is equivalent to the expectation value (\ref{length}). Hence, the above-mentioned observations are in agreement with those in
Refs.~\cite{burnett:PRA:1992} and ~\cite{Krause:PRA:1992}, where a background in the HHG
spectra of the dipole length was attributed to irreversible ionization at the end of the
pulse, and to the probability density near the edges of the integration box, respectively. One may select the spatial region near the core by employing the dipole acceleration \cite{burnett:PRA:1992}. In the results that follow, we will then employ Eq.~(\ref{accel1B}) for individual Bohmian trajectories or ensembles thereof. As a benchmark, we will take the expectation value (\ref{accel1}) computed directly from the TDSE.

\begin{figure}
 \begin{center}
 \includegraphics[width=8.5cm]{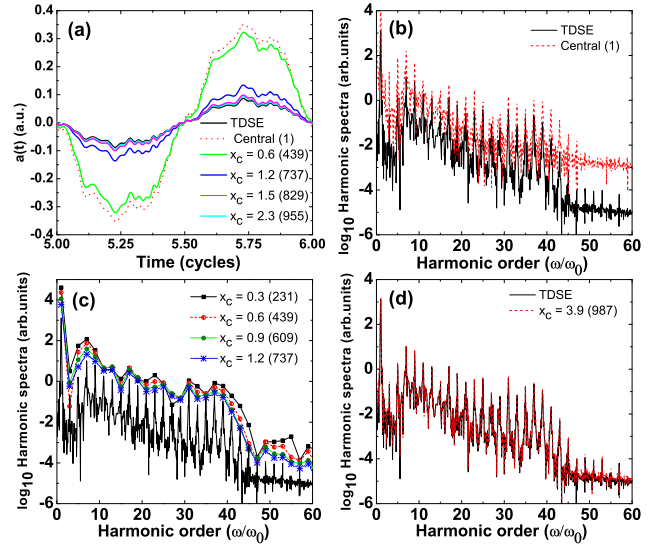}
 \end{center}
 \caption{\label{fig4}(Color online)
  Dipole acceleration obtained from a series of sets of Bohmian
  trajectories randomly distributed within the interval $[-x_c,x_c]$
  for  $E_0=0.075$ a.u. within one cycle of the pulse [panel (a)], together with harmonic spectra computed for the dipole acceleration [panels (b) to (d)]. The power spectra obtained for the central trajectory [Fig.~\ref{fig4}(b)], intermediate ensembles [Fig.~\ref{fig4}(c)] and the minimal ensemble for which a quantitative agreement with the TDSE occurs [Fig.~\ref{fig4}(d)] are displayed. In all cases, the black solid line represents the result obtained
  from the TDSE and the other lines represent the results from different ensembles of Bohmian trajectories.
  The different values of $x_c$ used in our simulations are labeled
  with different types of colors/line-styles (in parenthesis, we provide the total number of
  Bohmian trajectories used in each case). For the high-plateau harmonics, a quantitative agreement with the TDSE occurs already if 75\% of the overall probability
density is considered ($x_c=1.2$ a.u.), while, for the  lower
harmonics, 90\%
of the total probability density must be taken ($x_c=1.8$ a.u., not showing here). The remaining parameters are the same as in Fig.~\ref{fig2}.
  }
\end{figure}
We will now investigate which regions within the core must
be included for a quantitative agreement with the TDSE to be reached. This will be done by gradually varying the range of
$x_c$ of the random-number distribution
\footnote{The total interval has been chosen such that the initial
probability density,
$|\Psi(x,0)|^2$, is recovered up to over 99.8\%. This yields fourteen intervals at each side of
the origin, $x=0$, and around $N \approx 1,\!000$ random numbers in
total. Each of these numbers provides an initial
condition for a Bohmian trajectory.}. These results are displayed in
Fig.~\ref{fig4}(a) over a single cycle of the driving field, for the time-dependent dipole acceleration computed
using Eq.~(\ref{accel1B}). Overall, the
time dependence of all the distributions considered is very similar.
What is affected is the amplitude of the time-dependent
acceleration. If only the acceleration along the central trajectory
is taken (red lines in the figure), this amplitude is up
to five times larger than that determined by the acceleration
computed with the TDSE. The main effect of increasing the range of
the integration around this central trajectory is to decrease this
amplitude.

In Figs.~\ref{fig4}(b) to \ref{fig4}(d), we display the
high-harmonic spectra. Due to the above-mentioned difference in
amplitude, the harmonics in the spectrum from the central trajectory
are several orders of magnitude higher than those from the TDSE.
 As the spatial range is increased, a
 quantitative agreement is reached fairly quickly for the harmonics in the cutoff region, while for the below-threshold and low-plateau harmonics, larger ensembles of trajectories are needed.
Physically, this is consistent with what is known from strong-field
models, i.e., that the high-plateau and cutoff harmonics can be
reasonably modeled by the SFA, while the lower harmonics are much
more influenced by the internal structure of the system, such as excited states.

Nevertheless, even for the cutoff region an ensemble of Bohmian trajectories must also be taken if a quantitative agreement with the TDSE is to be achieved. This can be attributed to the fact that, quantum mechanically, there is a certain spread in the initial position of the electron, which must be considered. A quantitative agreement with the TDSE near the cutoff has also been obtained for an improved SFA model \cite{Plaja_2007}, in which the steepest descent method has not been performed and the continuum-to-continuum transitions have been incorporated. Physically, these modifications have introduced a spread in position and momentum. In the language of the TSM, this implies that the electron is no longer required to return exactly to the site of its release or leave with vanishing momentum.
\subsection{Time-frequency maps and the phase of the wavefunction}
\label{timefrequency}

\begin{figure*}
\begin{center}
 \includegraphics[width=15cm]{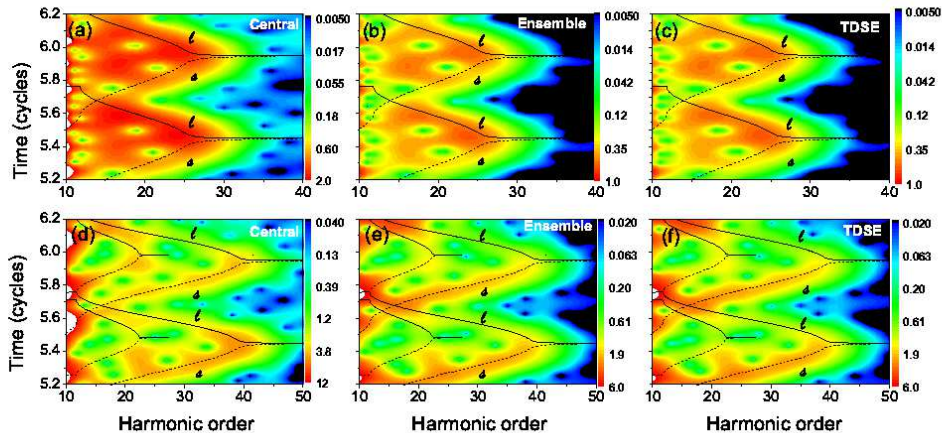}
\vspace*{-4cm} \caption{(Color online) Time frequency maps as
functions of the harmonic order computed using the Gabor transform
(\ref{Gabor}) of the dipole acceleration for trapezoidal fields of
frequency $\omega=0.057$ a.u. and intensities $I=8.7 \times
10^{13}\mathrm{W}/\mathrm{cm}^2$ [panels (a) to (c)], and $I=1.97
\times 10^{14}\mathrm{W}/\mathrm{cm}^2$ [panels (d) to (f)]. In
panels (a) and (d), only the central Bohmian trajectory is
considered in order to compute the acceleration $\overline{a}_B(t)$,
while in panels (b) and (e) an ensemble of 955 Bohmian trajectories
is taken. In panels (c) and (f), we display the results obtained
from the TDSE employing the expectation value of the dipole
acceleration [Eq.~(\ref{accel1})]. The black lines in the figure
indicate the real parts of the return times $t$ obtained from the
strong-field approximation, according to Eqs.~(\ref{t'saddle}),
(\ref{tsaddle}). In such equations, we have taken the ionization
potential to be the absolute value of the ground-state binding
energy ($|\epsilon_0|=0.66995$ a.u.). The dashed lines and the
letter $s$ indicate the short orbits, while the solid lines and the
letter $l$ indicate the long orbits. In the lower panels, there is a
second set of arches at lower energy. These arches corresponds to a
longer pair of return times.  The time is displayed in units of the
field cycle, and the yield has been plotted in a logarithmic scale. Throughout, the yield has been multiplied by 100.}
\label{Gab_Fig1}
\end{center}
\end{figure*}

In Fig.~\ref{Gab_Fig1}, we display time-frequency maps obtained for the dipole acceleration considering the central Bohmian trajectory, the ensemble average (\ref{accel1B}) and, for consistency, the expectation value of the dipole acceleration from the TDSE (left, middle and right panels, respectively). For comparison, superimposed to the time-frequency maps, we also plot the real parts of the return times $t$ obtained from the HHG transition amplitude in the SFA using the steepest descent method \cite{lewenstein:PRA:1994}.
 The times $t$, together with the ionization times $t^{\prime}$, are the solutions of the saddle-point equations
\begin{equation}  \label{t'saddle}
\frac{[\mathbf{p}(t,t^{\prime})+\mathbf{A}(t^{\prime})]^2}{2}+|\epsilon_0|=0,
\end{equation}
and
\begin{equation}  \label{tsaddle}
\frac{[\mathbf{p}(t,t^{\prime})+%
\mathbf{A}(t)]^2}{2}+|\epsilon_0|-\omega=0,
\end{equation}
where
\begin{equation}  \label{psaddle}
\mathbf{p}(t, t^{\prime})=-\frac{1}{(t-t^{\prime})}\int^t_{t^{\prime
}} d\tau\mathbf{A}(\tau)
\end{equation}
is the intermediate momentum of the released electron.
Eqs.~(\ref{t'saddle}) and Eq.~(\ref{tsaddle}) gives the conservation of energy at the instant of tunneling and recombination, respectively. At the recombination time $t$, a high-order harmonic of
frequency $\omega$ is generated \cite{lewenstein:PRA:1994}. The trajectories obtained from these solutions
are widely known as ``quantum orbits''
\cite{Salieres_2001}. Their real
parts are related to the classical trajectories of an electron in a
laser field, and form arch-like structures in the time-frequency maps, that merge at the cutoff. The upper and lower parts of each curve are related to the long and the short orbits, for which the electron returns after or before the field crossing \cite{Antoine_1996}.
In all panels of Fig.~\ref{Gab_Fig1}, such arch-like structures are present, both for the ensemble average (\ref{accel1B}) and for the acceleration along the central trajectory. These structures are particularly clear if the driving-field intensity is such that the system is in the tunneling regime [panel (d)]. In this case, the effect of taking an ensemble is that the relative contribution of the long orbit, with regard to that of the short orbit, is suppressed [see panel (e)].

This is consistent with other results in the literature, in which different weights between the signals related to the short and long trajectories have been observed in time-frequency \cite{Chirila_2010,Perez_2010,Wu2013} and intensity-reciprocal intensity maps \cite{Gaarde_2002}. In all cases, there seems to be a strong correlation between the over-enhancement of the long trajectory and the core region being strongly localized.
An enhancement of the long orbit occurs for short-range potentials, if compared with the their long-range counterparts \cite{Chirila_2010,Wu2013}, or in the SFA, if compared with the TDSE \cite{Gaarde_2002,Perez_2010}.
This over-enhancement is particularly extreme if the steepest descent method is employed in the SFA \cite{Gaarde_2002}.

Physically, this may be understood as follows. First, the long orbit is associated to a higher degree of wave-packet spreading, which means that it will be more sensitive with regard to spatial propagation, and to the uncertainty introduced by the initial momentum and position spreads. If the steepest descent method is used in the SFA, these initial spreads are neglected, and the long orbit is over-enhanced. Second, for a short-range potential, the initial wavepacket is highly localized, the effective barrier is very steep and there is no Rydberg series. This implies that (i) there will not be a strong mixing between excited states and the continuum in the presence of the field; (ii) it will be more difficult for an electron to reach the continuum along the short orbit, whose start times are farther away from the field peak. This will lead to over-enhanced contributions of the long orbits in the time-frequency maps.

In the present framework, using only the central Bohmian trajectory or small ensembles around it means that a high degree of localization is being imposed upon the core.  This has similarities with the SFA for which it is approximated by a single point at $x=0$, and with short-range potentials, for which there are no weakly bound states. In order to reproduce features that depend on the internal structure of the core, and to account for its spatial extension, one must employ larger ensembles of Bohmian trajectories (see our discussion of Fig.~\ref{fig4}). The less localized the initial wave packet is, the more trajectories one must consider.

For lower intensities, in the multiphoton regime, these features are more blurred, as shown in panel (a), in agreement with the results in Ref.~\cite{Chirila_2010}. This blurring is related to the appreciable acceleration of the electronic wavepacket outside the time intervals predicted by the TSM. These features are present in the SFA if the steepest descent method is not used \cite{Perez_2007}.

The above-stated results confirm that the phase of the wavefunction
mimics the behavior of a classical electron leaving and returning to
the core. Hence, one expects that, if this phase is somehow altered
or degraded, this will be transmitted non-locally to the central
Bohmian trajectory. Evidence for this non-locality has been provided
in our previous publication \cite{Wu2013}, using short- and
long-range potentials. Below we will investigate in more detail how
this phase builds up in configuration space. For that purpose, we
systematically disrupt the propagation the wavefunction outside the
core region, but avoid absorption of the probability flow within the
region for which the dipole acceleration is significant.

\begin{figure}
\hspace*{-0.5cm}\includegraphics[width=9.5cm]{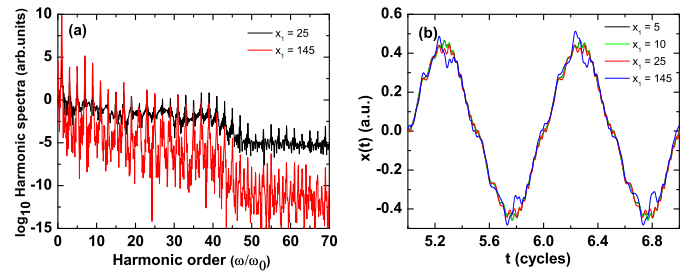}
\caption{(Color online) Panel (a): Harmonic spectra computed from the central Bohmian trajectory placing the absorber at $x_1=145$ a.u. and $x_1=25$ a.u. Panel (b): central Bohmian trajectory as a function of time over two cycles of the driving fields, for absorbers placed at varying distances $x_1$.  The external field has frequency $\omega=0.057$ a.u. and intensity $I=1.97 \times 10^{14}\mathrm{W}/\mathrm{cm}^2$ ($E_0=0.075$ a.u.), and the atomic parameters are the same as in the remaining figures. For the parameters employed in this figure, the classical excursion amplitude is $\alpha=23$ a.u.}
\label{mask}
\end{figure}

 This is performed by moving the absorber from near the boundaries of the integration box to ranges of the coordinate $x$ within the excursion amplitude $\alpha=E_0/\omega^2_0$ of a classical electron in the field. The lowest absolute value of $x_1$ in Eq.~(\ref{absorber}) was around $5$ a.u.(according to Sec.~\ref{ensemble}, quantitative agreement with the TDSE are expected for orbits starting in the interval $x\in[-2.3,2.3]$). For clarity, in Fig.~\ref{mask}, we show that, already for an absorber placed slightly beyond $\alpha$, there is substantial degradation in the spectra from the central Bohmian trajectory, but that the plateau and the cutoff are still present [panel (a)]. In contrast, the central Bohmian trajectories, as functions of time, look very similar [panel (b)].
\begin{figure}
\noindent
\hspace*{-0.5cm}\includegraphics[width=9cm]{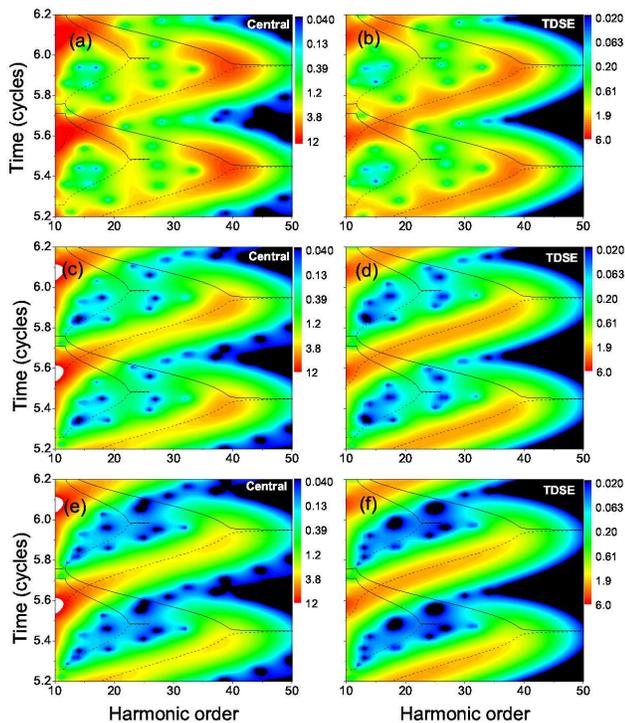}
\vspace*{-2cm}
 \caption{(Color online) Time frequency maps as
functions of the harmonic order computed using the Gabor transform
(\ref{Gabor}) for a trapezoidal field of frequency $\omega=0.057$
a.u. and intensity $I=1.97 \times 10^{14}\mathrm{W}/\mathrm{cm}^2$.
For the upper, middle and lower panels we have placed the absorber
given by Eq.~(\ref{absorber}) at $x_1=25$ a.u. [panels (a) and (b)],
$x_1=10$ a.u. [panels (c) and (d)] and $x_1=5$ a.u. [panels (e) and (f)], respectively.  In the left and right panels we have
computed the Gabor transforms of the dipole acceleration along the
central Bohmian trajectory [panels (a), (c) and (e)], and of its expectation value obtained from the TDSE [panels (b), (d) and
(f)], respectively. The black lines in the figure indicate the real
parts of the solutions of the saddle-point equations for the return
time $t$. The dashed lines correspond to the short orbits in a pair,
while the solid lines give the long orbits.  The yields in all panels have been multiplied by 100.} \label{Gab_Fig2}
\end{figure}

The time-frequency maps obtained in this way are presented in
Fig.~\ref{Gab_Fig2} for the higher intensity employed in
Fig.~\ref{Gab_Fig1}. In the figure, the patterns associated with the
longer orbits become increasingly suppressed and deteriorate as the
absorber is moved towards the core. For instance, if the absorber is
placed slightly beyond the classical excursion amplitude, the arch corresponding to the
longer SFA return times, which extends to around the 25th harmonic,
becomes fainter. This holds for the time-resolved spectra of the central trajectory [Fig.~\ref{Gab_Fig2}(a)]
and of the outcome of TDSE [Fig.~\ref{Gab_Fig2}(b)].
The same
behavior is observed for the feature associated with the long
trajectory belonging to the shortest pair. If the absorber is made
active slightly below the middle of the electron classical excursion
amplitude, the above-mentioned features are very weak, as shown in
the middle panels of Fig.~\ref{Gab_Fig2}. For an absorber starting
immediately after the core region ($x_1=5$~a.u.), all the features
in the time-frequency maps associated with the longer trajectories
are substantially reduced [Figs.~\ref{Gab_Fig2}(e) and (f)]. This
shows that the probability density that leaves the core region plays
a very important role in influencing the phase of the wavefunction. Throughout, the absorber is extremely smooth, so that the probability density-flow is disrupted but not eliminated and full absorbtion only occurs for $x=\pm l$. For a sharper absorber, the signal related to the short orbit would also be eliminated.
\section{Conclusions}
\label{conclusions}

In this work, we have applied Bohmian mechanics to assess the
influence of specific regions in configuration space on high-order
harmonic generation (HHG). We have employed ensembles around the central Bohmian trajectory, starting at $x(0)=0$, which, previously, has been found to yield clear spectra, with a plateau and a cutoff \cite{Wu2013}.  If the ensemble
average (\ref{dipoleB}) is taken, these features are masked by the Bohmian trajectories associated with the
probability-density flow far from the core.
Furthermore, in the present framework, the
three-step model 
manifests itself in the phase of the wavefunction. This phase mimics the behavior of an ensemble of classical electron trajectories leaving
and returning to its parent ion. Time-frequency analysis shows that this pattern is present in the
probability flow that remains at the core.

Nevertheless, the probability flow leaving and returning to the core does influence how this phase builds up. In fact, by deliberately placing absorbing boundaries in the spatial regions within the ranges of excursion amplitudes of a classical electron, but outside the core region, we have degraded the phase of the whole wavefunction. This has influenced both the HHG spectrum
and the time-frequency maps of the central Bohmian trajectory, and the degradation has occurred as it would be expected from the three-step model (TSM). In fact, the first sets of features to be eliminated from the time-frequency maps as the absorber increasingly approaches the core correspond to the trajectories with longer excursion times. These changes are transmitted nonlocally, as the probability density flow associated with the central Bohmian trajectory never leaves the core region. Since Bohmian trajectories are in fact ``slices" of the wavefunction, they will always depend on the behavior associated with the whole wavefunction, regardless of the positions occupied by the probability density in configuration space. Hence, if the wavefunction is altered far from the core, all the ``slices" will be different. One should note that the probability-density flow could have been altered in other ways, such as employing inhomogeneous media \cite{Ciappina_1_2012,Ciappina_2_2012}. However, moving boundaries allow the systematic deconstruction of the flow in the desired spatial range.

Our results also show that a spatially extended core is paramount for obtaining a good agreement with the  TDSE, even if this spatial extension is small. This holds not only for the low-plateau and below-threshold harmonics, but also for the cutoff energy region. This implies that an initial wave-packet spread in position space is necessary for an accurate description of the problem. From a Bohmian-trajectory perspective, this spread manifests itself as an uncertainty in the initial conditions $x_i(0)$ \cite{bohm:PR:1952-1}. This is consistent with recent studies in which the strong-field approximation (SFA) has been improved to include this spread \cite{Perez_2007,Plaja_2007}. Its fully quantum mechanical version, without resorting to the steepest descent method, has shown a very good agreement with the TDSE in the cutoff region.  In the time-frequency maps, neglect of the initial position spread causes an over-enhancement in the signal related to the long SFA trajectory \cite{Gaarde_2002,Perez_2010}. We have identified a similar effect in our computations, if only the central Bohmian trajectory is considered. An important issue here is non-locality. In the full quantum
mechanical version of the SFA, the time-dependent wave
function is described as a coherent superposition of the ground
state and the continuum, which is approximated by field-dressed plane waves. In this case, the continuum is also non-local. Due to the approximations involved, however, the phase picked up by the time-dependent wavefunction is different from that of the full TDSE. Bohmian trajectories consider the full phase, but may restrict the initial spread in position space.

Finally, we would like to elaborate on the advantages and disadvantages of Bohmian trajectories. On the one hand, since Bohmian trajectories contain all phase information related to the full time-dependent Hamiltonian, both the driving field and the binding potential are accounted for.  They may thus be used to probe the time-dependent wavefunction in specific regions in configuration space, and phenomena associated with this wavefunction, such as the example provided in this article. They can also be employed for visualization purposes, in order to access how the time-dependent probability density flow behaves in particular spatial regions, e.g., near the core or far from it. On the other hand, these trajectories are highly non-local entities. Therefore, if one is interested in information related to the phase of the wavefunction, and not the probability-density flow, one needs to employ additional resources, such as time-frequency maps, or quasiprobabilities in phase space.

Another advantage is that, in principle, Bohmian trajectories may be used for reconstructing the time-dependent probability flow, or parts thereof. In practice, however, a huge obstacle to this reconstruction is that, according to the present scheme, it is first necessary to solve the full TDSE in order to obtain these trajectories. This is particularly problematic for systems with many degrees of freedom. There is, however, a real effort in order to overcome this obstacle, and construct Bohmian trajectories without the need for the TDSE solution \cite{Schleich_2013}.
\acknowledgements
This work was supported by the UK EPSRC (grant
EP/J019240/1), and the CSC/BIS.  We thank X.
Lai, C. Zagoya and in particular A. Fring for useful comments and discussions.
We are also grateful to A. Sanz for his collaboration in the initial stages of this project.

\end{document}